\documentclass{aastex}
\usepackage{spr-astr-addons}
\usepackage{url}

\RequirePackage{color}

\begin{document}

\title{Self-similar solutions of viscous and resistive ADAFs with thermal conduction}
\shorttitle{Viscous and resistive ADAF with thermal conduction}
\shortauthors{K. Faghei}

\author{\\ Kazem Faghei
}
\affil{School of Physics,
Damghan University, Damghan, Iran\\
e-mail: kfaghei@du.ac.ir \\
\\
Received: 14 October 2011 / Accepted: 30 November 2011 }


\begin{abstract}
We have studied the effects of thermal conduction on the structure of viscous and resistive advection-dominated accretion 
flows (ADAFs). The importance of thermal conduction on hot accretion flow is confirmed by observations of hot gas that
 surrounds Sgr A$^*$ and a few other nearby galactic nuclei. In this research, thermal conduction is studied by a saturated
 form of it, as is appropriated for weakly-collisional systems. It is assumed the viscosity and the magnetic diffusivity
 are due to turbulence and dissipation in the flow. The viscosity also is due to angular momentum transport. Here, the
 magnetic diffusivity and the kinematic viscosity are not constant and vary by position and $\alpha$-prescription is used 
for them. The govern equations on system have been solved by the steady self-similar method. The solutions show the radial 
velocity is highly subsonic and the rotational velocity behaves sub-Keplerian. The rotational velocity for a specific value
 of the thermal conduction coefficient becomes zero. This amount of conductivity strongly  depends on magnetic pressure
 fraction, magnetic Prandtl number, and viscosity parameter. Comparison of energy transport by thermal conduction
 with the other energy mechanisms implies that thermal conduction can be a significant energy mechanism in resistive and
 magnetized ADAFs. This property is confirmed by non-ideal magnetohydrodynamics (MHD) simulations.
\end{abstract}
\vspace{0.5cm}
\vspace{0.5cm}
\section{Introduction}
The observational features in active galactic nuclei (AGN) and X-ray binaries can be successfully explained by the standard 
geometrically thin, optically thick accretion disc model (Shakura \& Sunyaev 1973).  The motion of the matter in the
 standard thin model of the accretion disc is approximately Keplerian, and the energy released in the accreting gas is
 radiated away locally. In the past two decades, another type of accretion flow has been considered that the energy released 
due to heating processes in the flow may be trapped within accreting gas. As, only the small
 fraction of the energy released in the accretion flow is radiated away due to inefficient cooling, and most of the energy 
is stored in the accretion flow and advected to the central object. This type of accretion flow is called as advection-dominated 
accretion flow (ADAF). The models of ADAF have been studied by a number of researchers (e.g.
 Ichimaru 1977; Rees et al. 1982; Narayan \& Yi 1994; Abramowicz et al. 1995; Blandford \& Begelman 1999; Ogilvie
 1999).

The observations of black holes confirm existence of hot accretion flow that contrasted with classical cold and thin
 accretion disc model (Shakura \& Sunyaev 1973). Hot accretion flows can be seen in supermassive black
 holes of galactic nuclei and during quiescent of accretion onto stellar-mass black holes in X-ray transients (e.g., Lasota
 et al. 1996; Esin et al. 1997, 2001; Narayan et al. 1998a; Menou et al. 1999;  Di Matteo et al. 2000; see Narayan et al.
 1998b; Melia \& Falcke 2001; Narayan 2002; Narayan \& Quataert 2005 for reviews). \textit{Chandra} observations provide
 constraints on the density and temperature of gas at or near the the Bondi capture radius in Sgr A$^*$ and several
  nearby galactic nuclei (Loewenstein et al. 2001; Baganoff et al.
 2003; Di Matteo et al. 2003; Ho et al. 2003). Tanaka \& Menou (2006) exploited these constraints to calculate mean free path of the observed gas. 
They suggested  accretion in such systems are under weakly collisional condition. Moreover, they suggested thermal conduction as a
 possible mechanism by which the sufficient extra heating is provided in hot accretion flows. In following, Johnson \&
 Quataert (2007) studied the effects of electron thermal conduction on the properties of a spherical hot accretion flows.
 Their model is applicable for Sgr A$^*$ in the Galactic centre. Because, electron heat conduction is important for low
 accretion rate systems. They also found a supervirial temperature in the presence of thermal conduction. Similar to Tanaka \& Menou (2006), 
they assumed a steady state model, but they solved their equations numerically. Abbassi et al. (2008) presented a
 set of self-similar solutions for ADAFs with a toroidal magnetic field in which the saturated thermal conduction has a
 important role in the energy transport in the radial direction.  

Since the observational evidences and magnetohydrodynamics (MHD) simulations have expressed the toroidal magnetic field and
 the magnetic diffusivity are important in accretion flows (see Faghei 2011a and references therein),  Faghei (2011a)
 examined the self-similar solutions of viscous and resistive ADAFs in the presence of a toroidal magnetic field. But, he
 did not consider the effects of thermal conduction in his model. While, the recent studies of resistive accretion flows
 have represented that thermal conduction can play an importance role in such systems (e.g. Sharma et al. 2008; Ghanbari et
 al. 2009). Sharma et al. (2008) studied the effects of thermal conduction on magnetized spherical accretion flows using
 global axisymmetric MHD simulations. In their model, when the magnetic energy density becomes comparable to the
 gravitational potential energy density, the plasma due to resistivity is heated to roughly the virial temperature, the
 mean inflow becomes highly subsonic, and most of the energy released  by accretion is transported to large radii by
 thermal conduction and the accretion rate became much smaller than Bondi accretion rate.  Moreover, they found that for a
 larger values of conductive parameter, energy transport through thermal conduction becomes the dominant energy transport
 mechanism at small radii. Ghanbari et al. (2009) studied a two-dimensional advective accretion disc bathed in the poloidal
 magnetic field of a central accretor in the presence of thermal conduction. They did not consider toroidal component of
 magnetic field and for simplicity assumed the resistivity as a constant. They studied induction equation of magnetic field
 in a steady state that is not according to anti-dynamo theorem (e.g. Cowling 1981) and is useful only in particular
 systems where the magnetic dissipation time is much longer than the age of the system. On the other hand, this assumption
 implies that the flow is in balance between escape and creation of the magnetic field.
  
In this paper, we adopt the presented solutions by Tanaka \& Menou (2006) and Faghei (2011a). Thus, we will investigate the
 influences of thermal conduction on a viscous and resistive ADAF in the presence of a toroidal magnetic field. Moreover,
 it is assumed that magnetic diffusivity in the present model is not constant, and escaping and creating of magnetic field 
are unbalanced. From some aspects will be shown that the present model is in according with the observations and the 
resistive MHD simulations. The paper is organized as follows. In section 2, the basic equations of constructing a model for
 quasi-spherical magnetized advection dominated accretion flow with thermal conduction will be defined. In section 3, self-
similar method for solving equations which govern the behaviour of the accreting gas was utilized. The summary of the model
 will appear in section 4.  

\section{Basic Equations}
We suppose a rotating and accreting gas around a Schwarzschild  black hole of mass $M$. The flow is assumed to be in
 advection dominated stage, where viscous and resistive heating are balanced by the advection cooling and thermal
 conduction. We use a spherical coordinate ($r$,  $\theta$, $\phi$) centred on the accreting object. Furthermore, the flow 
is assumed to be steady and axisymmetric ($\partial_t=\partial_\phi=0$), and the equations will be considered in the
 equatorial plane, $\theta=\pi/2$. Thus, all flow variables are a function of only $r$. For the sake of simplicity, the
 general relativistic effects are ignored and Newtonian gravity is used. The magnetic field in the present model has only a
 toroidal component. Under assumptions, the model is described by the following equations:

The continuity equation with mass loss is
\begin{equation}
\frac{1}{r^{2}}\frac{d}{d r}(r^{2}\rho v_{r})=\dot{\rho},
\end{equation}
where $\rho$ is density, $v_r$ is the radial infall velocity, and $\dot\rho$ the mass-loss rate per unit volume. 

The radial equation of momentum is
\begin{equation}
v_{r}\frac{d
v_{r}}{d r}
  =r (\Omega^{2}-\Omega^{2}_K)-\frac{1}{\rho}\frac{d }{d r}(\rho c^2_s)
     -\frac{c^2_A}{r}-\frac{1}
     {2 \rho}\frac{d }{d r}(\rho c^2_A),
\end{equation}
 where $\Omega$ is the angular velocity of the flow, $\Omega_K=\sqrt{G M/ r^3}$ is the Keplerian angular velocity, $c_s$
 the isothermal sound speed, which is defined as $c_s^2=p_{gas}/\rho$, $p_{gas}$ being the gas pressure, and $c_A$ is
 Alfven speed, which is defined as $c_A^2 =B_{\varphi}^2/4\pi\rho=2 p_{mag}/\rho$, $p_{mag}$ being the magnetic pressure.
The angular momentum transfer equation is
\begin{equation}
\rho v_{r}\frac{d}{d r}(r^{2}\Omega)=
  \frac{1}{r^{2}}\frac{d}{d r}\left[\nu\rho r^{4}\frac{d \Omega}{\partial r}\right],
\end{equation}
where the right-hand side of above equation describes the effects of viscous torques due to shear ($\nu$, here, is 
kinematic coefficient of viscosity). As noted in the introduction, we assume both of the kinematic coefficient of viscosity
 and the magnetic diffusivity due to turbulence in the accretion flow. Thus, it is reasonable to use these parameters in 
analogy to the $\alpha$-prescription of Shakura \& Sunyaev (1973) for the turbulent,
\begin{equation}
   \nu = P_m \eta = \alpha \frac{c_s^2}{\Omega_{K}},
\end{equation}
where $P_m$ is the magnetic Prandtl number, which is assumed a constant of order of unity, $\eta$ is the magnetic
 diffusivity, and $\alpha$ is a free parameter less than unity. 

The energy equation becomes
\begin{eqnarray}
 \nonumber \frac{v_r}{\gamma-1}\frac{d}{d r}(\rho c^2_s)+
\frac{\gamma}{\gamma-1}\frac{ \rho c^2_s}{r^2}\frac{d}{d r}\left( r^2 v_r\right)= \\
Q_{diss}-Q_{rad}+Q_{cond},
\end{eqnarray}
where $Q_{diss}=Q_{vis}+Q_{resis}$ is the dissipation rate by viscosity $Q_{vis}$ and resistivity $Q_{resis}$,
$Q_{rad}$ represents the energy loss through radiative cooling, and $Q_{cond}$ is the energy transported by thermal
 conduction. For the right-hand side of the energy equation, we can write
 \begin{equation}
   Q_{adv}=Q_{diss}-Q_{rad}+Q_{cond}
\end{equation}
where $Q_{adv}$ is the advective transport of energy. We exploit the advection factor, $f=1-Q_{rad}/Q_{diss}$, that 
describes the fraction of the dissipation energy which is stored in the accretion flow and advected into the central object
 rather than being radiated away. In general, the advection factor depends on the details of the heating and radiative
 cooling mechanisms and will vary by position (e.g. Watari 2006, 2007; Sinha et al. 2009). Here, we assume a constant $f$
 for simplicity. Clearly, the flow in the case of $f = 1$ is in the extreme limit of no radiative cooling and in the limit
 of efficient radiative cooling, we have $f = 0$.

As mentioned, the inner regions of hot accretion flows are collisionless and the electron mean free path due to 
Coulomb collision is larger than the radius. This property is described as \textit{saturation} (Cowie \& McKee 1977). The
 traditional equation of heat flux due to thermal conduction, $F_{cond}=-\kappa \nabla T$  which $\kappa$ being the thermal
 conduction coefficient, is not valid for such systems. Because this equation is suitable for the collisional plasma, in
 which mean free path of electron energy exchange is smaller than temperature scale height. Cowie \& McKee (1977) derived
 the heat flux for the collisionless plasma as 
\begin{equation}
   F_{sat}=5 \phi_s \rho c_s^3=5 \phi_s p \sqrt{\frac{p}{\rho}},
\end{equation}
where $\phi_s$ is a factor is less than unity and is called as saturation constant. Now, the viscous and resistive heating
 rates and the energy transport by thermal conduction are expressed as
\begin{equation}
   Q_{vis}=\nu\rho r^2 \left(\frac{\partial \Omega}{\partial  r}\right)^2
\end{equation}
\begin{equation}
   Q_{resis}=\frac{\eta}{4\pi} {\bf J}^2
\end{equation}
\begin{equation}
   Q_{cond}=- \left|\frac{1}{r^2}\frac{\partial}{\partial r} \left(r^2 F_{sat}\right)\right|
\end{equation}
where ${\mathbf J}=\nabla\times {\mathbf B}$ is the current density, ${\mathbf B}$ being the magnetic field. We
 used a minus sign in equation (10). Because we want thermal conduction as energy transport mechanism outward. On the other
 hand, heat conduction flux will be behaved like a cooling mechanism in accretion flow. Finally, since we consider only
 the toroidal field, the induction equation with field escape can be written as
\begin{equation}
\frac{1}{r}\frac{d
}{d r}\left[r v_{r} B_{\varphi}-\eta\frac{d
}{d r}(r B_{\varphi})\right]=\dot{B}_{\varphi}.
\end{equation}
where $B_{\varphi}$ is the toroidal component of magnetic field and $\dot{B}_{\varphi}$ is the field escaping/creating rate 
due to a magnetic instability or  dynamo effect. This induction equation is rewritten as
\begin{eqnarray}
\nonumber\frac{1}{r}\frac{d
}{d r}\left[\sqrt{4\pi\rho c^2_A}\left(r v_{r} -\frac{\alpha}{4 \beta P_m }\frac{1}{ r \rho\Omega_K}
\frac{d}{d r}(r^2 \rho c^2_A ) \right)\right]=
\\
\dot{B}_{\varphi},
\end{eqnarray}
where $\beta$ is the degree of magnetic pressure to the gas pressure and can be defined by
\begin{equation}
\beta=\frac{p_{mag}}{p_{gas}}=\frac{1}{2} \left(\frac{c_A}{c_s}\right)^2.
\end{equation}
In this paper, we apply the steady self-similar methods to solve the system equations. Thus, this parameter will be 
constant throughout the disc. While, Khesali \& Faghei (2008, 2009) showed that it varies by position. In hot accretion 
flows, typical value of $\beta$ is in the range $0.01$-$1$ (e.g. De Villiers et al. 2003; Beckwith et al. 2008). Here, we 
will also consider the magnetically dominated case ($\beta > 1$). Because, when thermal instability happens in an ADAF, the 
MHD numerical simulations imply that the thermal pressure rapidly decreases while the magnetic pressure increases due to 
the conservation of magnetic flux (Machida et al. 2006). This will result in large $\beta$ and forms a magnetically 
dominated accretion flow (Bu et al. 2009). 

\section{Self-Similar Solutions}
\subsection{Analysis}
The self-similar method is useful to understand of the physics of accretion flows. This method is familiar due to its wide 
range of applications in many research fields of astrophysics. Self-similar solution, although constituting only a limited 
part of problem, is often useful to understand the basic behaviour of the system. Thus, in order to seek similarity 
solutions for the above equations, we seek solutions in the following form:
\begin{equation}
v_r(r)=-c_1 \alpha \sqrt{\frac{G M_*}{r}}
\end{equation}

\begin{equation}
\Omega(r)=c_2\sqrt{\frac{G M_*}{r^3}}
\end{equation}

\begin{equation}
c^2_s(r)=c_3\frac{G M_*}{r}
\end{equation}

\begin{equation}
c^2_A(r)=\frac{B^2_{\varphi}}{4\pi\rho}=2 \beta c_3\frac{G M_*}{r}
\end{equation}
where coefficients $c_1$, $c_2$, and $c_3$ are determined later. We assume a power-law relation for density
\begin{equation}
\rho(r)=\rho_0 r^s
\end{equation}
where $\rho_0$ and $s$ are constant. By using above self-similar quantities, the mass-loss rate and the field 
escaping/creating rate must have the following form:

\begin{equation}
\dot{\rho}(r)=\dot{\rho}_0 r^{s-3/2}
\end{equation} 

\begin{equation}
\dot{B}_{\varphi}(r)=\dot{B}_0 r^{\frac{s-4}{2}}
\end{equation}
where $\dot{\rho}_0$ and $\dot{B}_0$ are constant. 

Substituting the above solutions in the continuity, momentum, angular momentum, energy, and induction equations [(1)-(3), 
(5), and (11)], we can obtain the following relations:

\begin{equation}
\dot{\rho}_0=-\left(s+\frac{3}{2}\right) \alpha \rho_0 c_1 \sqrt{G M_*},   
\end{equation}

\begin{equation}
-\frac{1}{2}c^2_1 \alpha^2 = c^2_2 -1 - c_3 \left[s-1+\beta (1+s)\right],
\end{equation}

\begin{equation}
c_1=3 (s+2) c_3, 
\end{equation}
\begin{eqnarray}
\nonumber-\alpha c_1 \left[\frac{2(s-1)+3 \gamma}{\gamma-1}\right]=\alpha f \left[
\frac{9}{2} c^2_2 +\frac{\beta}{P_m} c_3 (1+s)^2
\right]\\
- 10 \phi_s c_3^{1/2} \left|s+\frac{1}{2}\right|, 
\end{eqnarray} 

\begin{equation}
\dot{B}_0=-\frac{\alpha s}{2} G M_* \sqrt{2\pi\rho_0\beta c_3}\left[
2 c_1  + \frac{c_3}{P_m}  (1+s) \right].  
\end{equation} 
Above equations express for $s=-3/2$, there is no mass loss, while for $s > -3/2$ mass loss (wind) exists. After algebraic 
manipulations, we obtain an algebraic equation for $c_3$:
\begin{eqnarray}
\nonumber{\frac {81}{8}}\,{\alpha}^{3} (s+2) ^{2}f{c_{{3}}}^{2}+
 c_{3} \Bigg  [ \frac{3\, \alpha (s+2)}{2} \times \frac{2(s-1)+3 \gamma}{\gamma-1} 
\\
\nonumber
-\frac{9 \alpha f}{4}\, \Bigg( \frac{2 \beta}{9 P_m} ( s+1) ^{2}
+ ( s+1) \beta+s-1 \Bigg)  \Bigg]
\\
 -5  \phi_{s} \left|s+\frac{1}{2}\right|
\sqrt {c_{{3}}}-\frac{9}{4}\,f\alpha=0,
\end{eqnarray}
and the rest of the physical variables are
\begin{equation}
\dot{\rho}_0=-3\alpha \rho_0 \sqrt{G M_*} (s+2) \left(s+\frac{3}{2} \right) c_3,
\end{equation}
\begin{equation}
c_1=3 c_3 (s+2),
\end{equation}
\begin{equation}
c_2^2=1-\frac{9  \alpha^2 }{2}(s+2)^2 c_3^2+c_3 \left[ (s+1) \beta + s-1 \right],
\end{equation}
\begin{eqnarray}
\nonumber \dot{B}_0=-3\alpha s G M_* (s+2)^{3/2} c_3^{3/2} \sqrt{6 \pi \rho_0 \beta} \times 
\\  \left[ 1+\frac{s+1}{6 P_m(s+2)} \right].
\end{eqnarray}
 
We can solve algebraic equation (26) numerically and clearly only real roots which correspond to positive $c_1$ are 
physically acceptable. Without thermal conduction, i.e. $\phi_s=0$, (26) and similarity solutions reduce to Faghei (2011a). 
But our main algebraic equation includes thermal conduction. 

\input{epsf}
\begin{figure}[t]
\begin{center}
\centerline
{ 
{\epsfxsize=4.3cm\epsffile{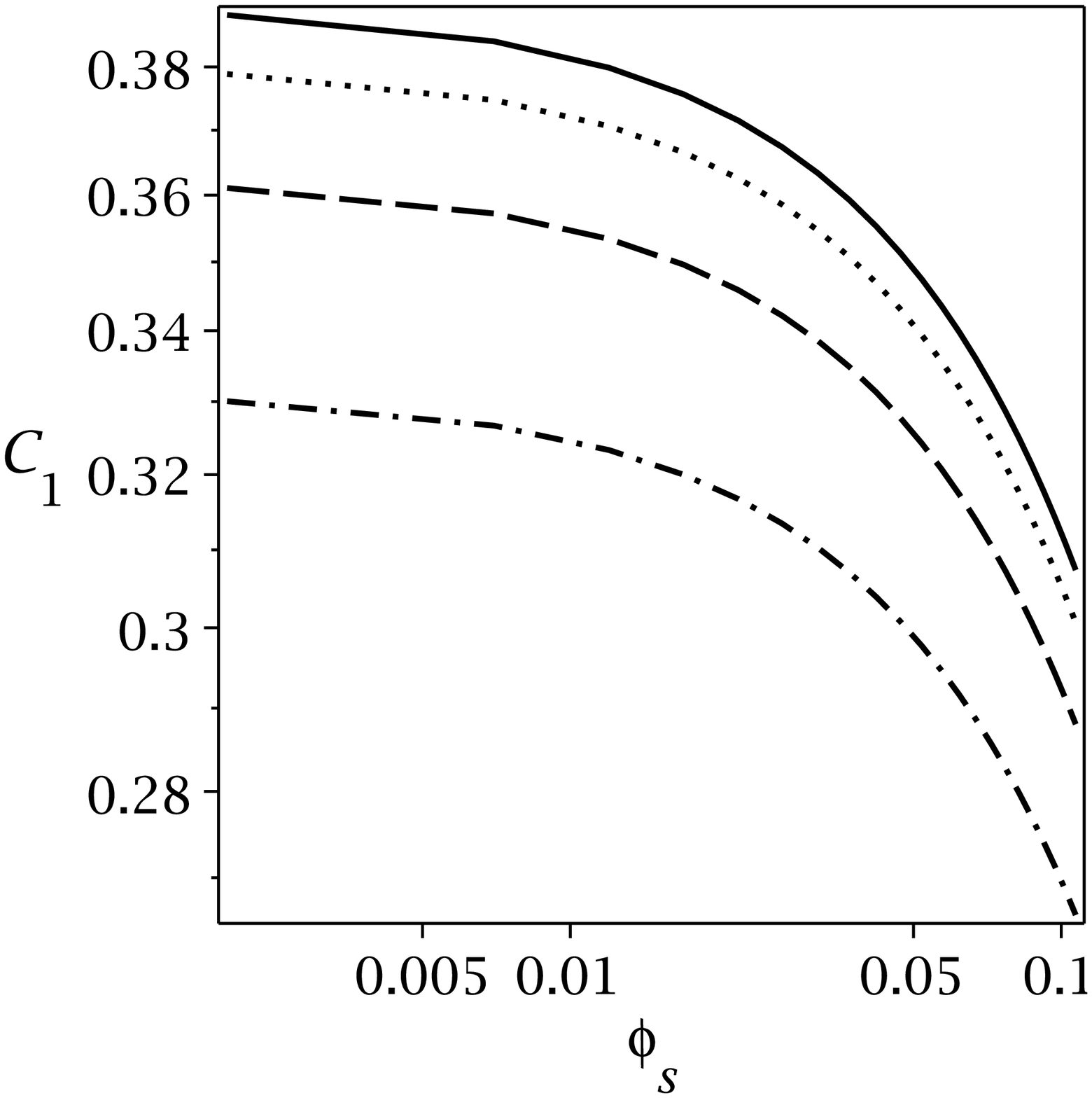}  }
{\epsfxsize=4.3cm\epsffile{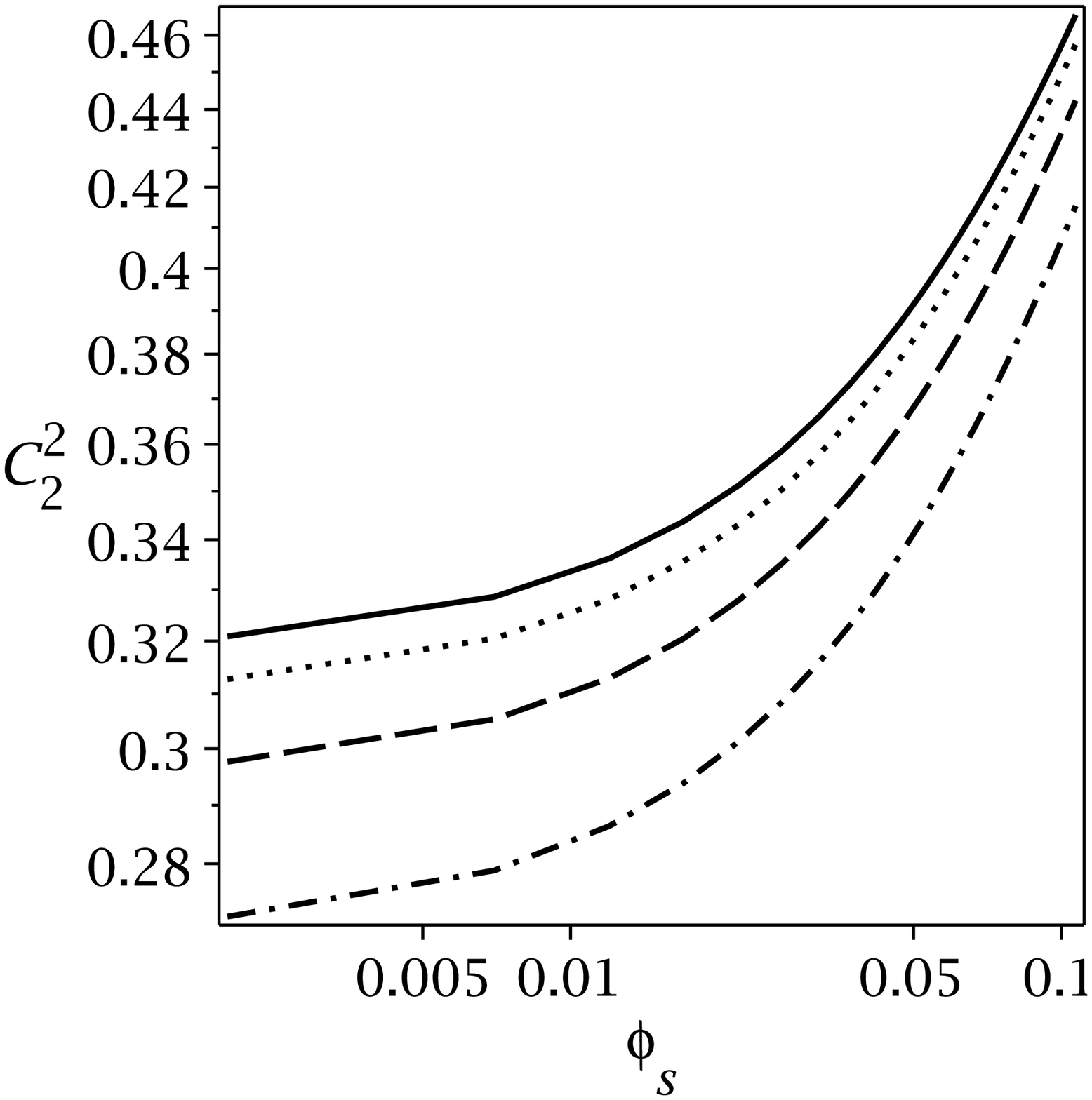} }
} 
\centerline
{ 
{\epsfxsize=4.3cm\epsffile{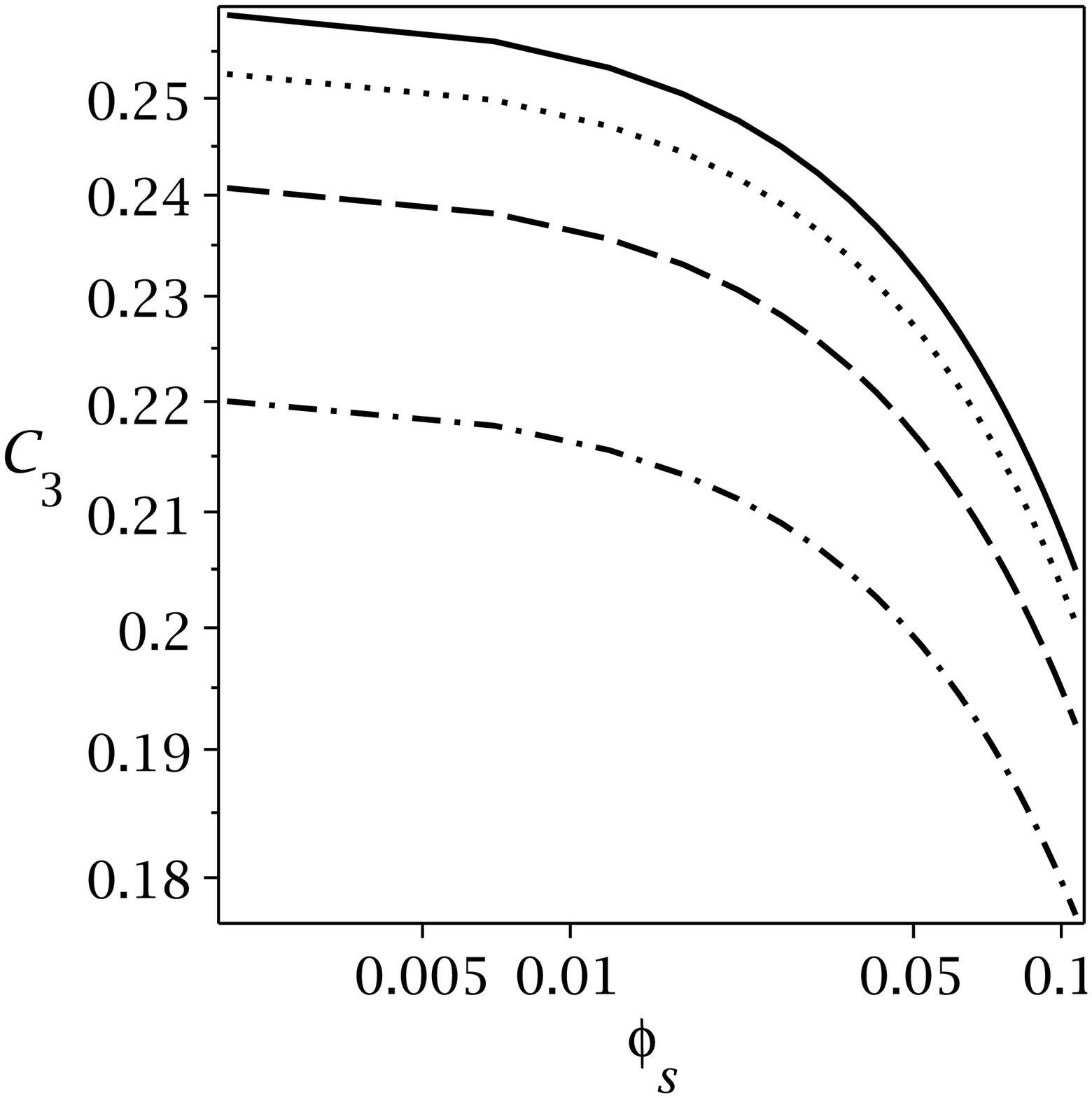}  }
{\epsfxsize=4.3cm\epsffile{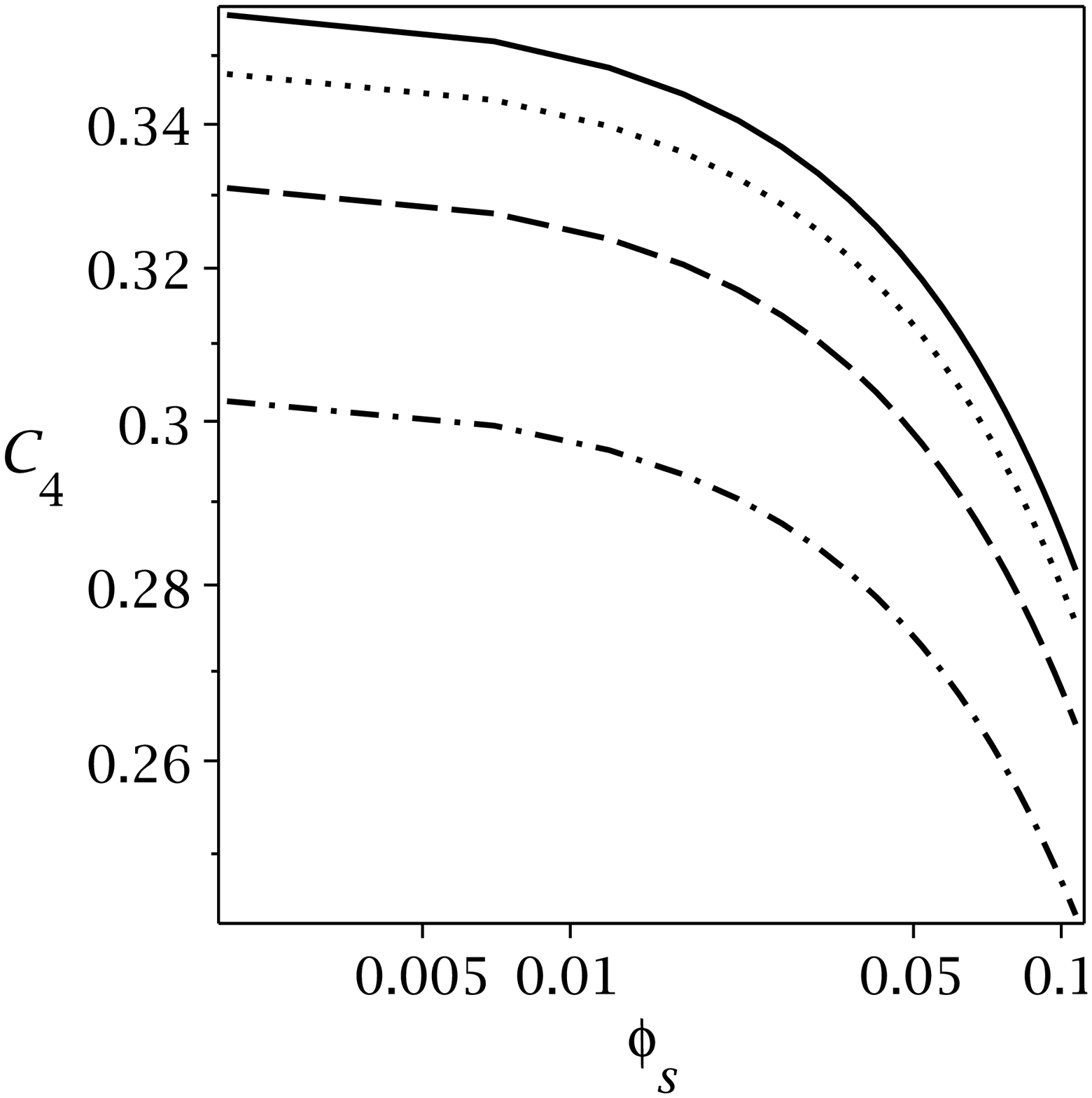}  }
}      
\end{center}

\begin{center}
\vspace{-0.5cm}
\caption{Physical quantities of the flow as a function of saturation constant for $\gamma=1.3$, $\alpha=0.5$, $f=1$, and 
$P_m=100$. Solid, dotted, dashed, and dot-dashed lines represent $\beta=0.1, 0.3, 0.7$, and $1.5$. }
\end{center}
\end{figure}

\subsection{Numerical Results}
Now, we consider the behaviour of solutions in the presence of thermal conduction. But in this paper, only case of no wind 
($s=-3/2$) is considered that $\dot\rho_0=0$ and $\dot{B}_{\varphi}\propto r^{-11/4}$. In addition to introduced 
coefficients, we also define a new parameter of $c_4$ that is the right-hand side of equation (24)
\begin{equation}
c_4=\alpha f \left[
\frac{9}{2} c^2_2 +\frac{\beta}{P_m} c_3 (1+s)^2
\right]
-10 \phi_s c_3^{1/2} \left|s+\frac{1}{2}\right|. 
\end{equation} 
Equations (5), (24), and (31) imply that the parameter $c_4$ is the advection transport of energy. The behaviour of 
coefficients $c_i$ as a function of $\phi_s$ are shown in Figures 1-3. Moreover, Figure 1 represents the profiles of 
physical quantities for several values of the magnetic pressure fraction, i. e. $\beta=0.1, 0.3, 0.7$, and 
$1.5$. The value of $\beta$ measures the strength of magnetic field, and a larger $\beta$ denotes a stronger magnetic 
field. Figure 2 represents the profiles of physical quantities for several values of magnetic Prandtl number, i. e. $P_m = 
\infty, 1, 2/3$, and $1/2$. The smaller values of $P_m$ denotes a stronger magnetic diffusivity, $\eta$. Figure 3 shows the 
profiles of physical quantities for several values of adiabatic index, i. e. $\gamma = 1.2, 1.25, 1.3$, and $1.35$.

\input{epsf}
\begin{figure}[t]
\begin{center}
\centerline
{ 
{\epsfxsize=4.3cm\epsffile{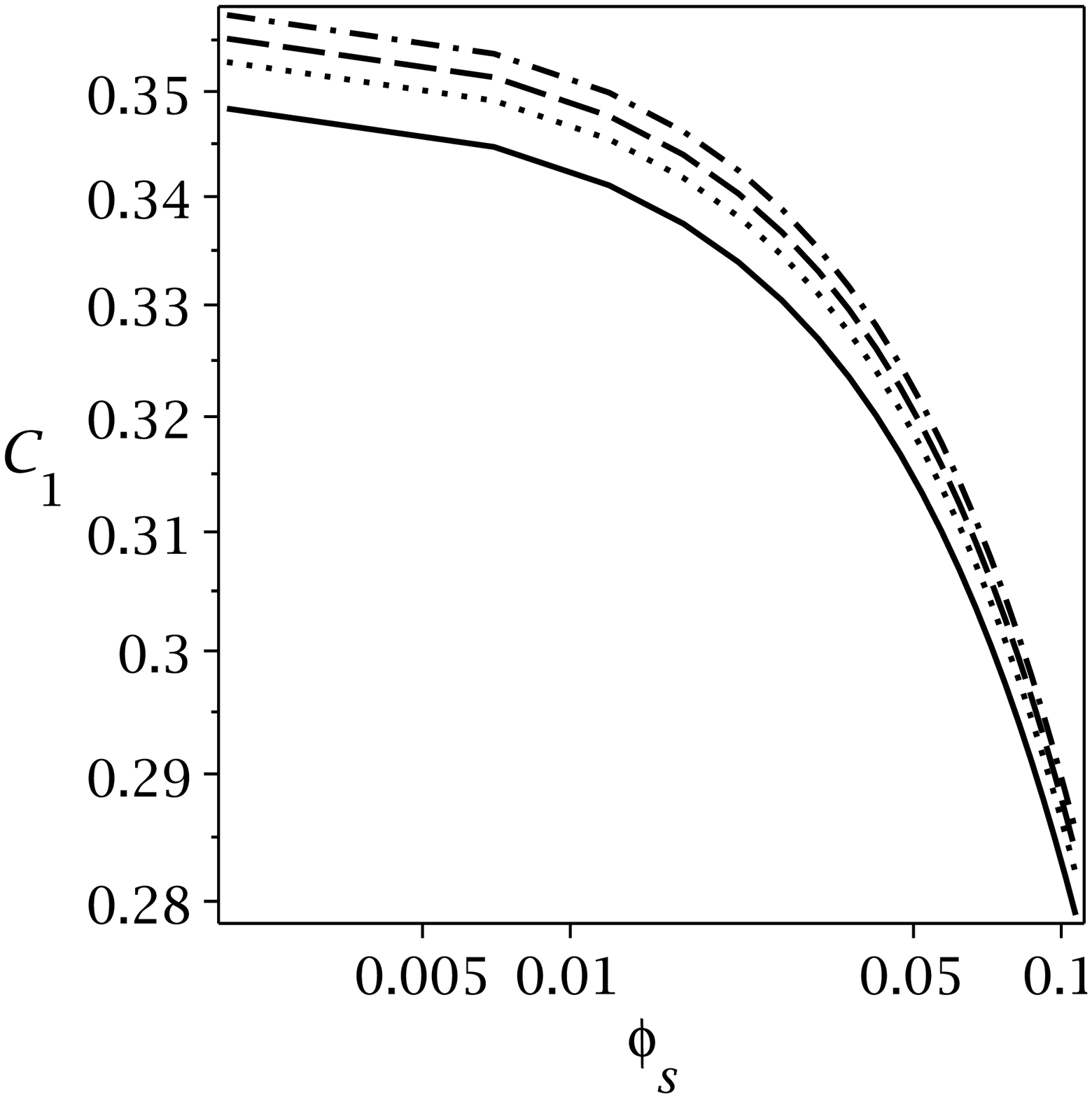}  }
{\epsfxsize=4.3cm\epsffile{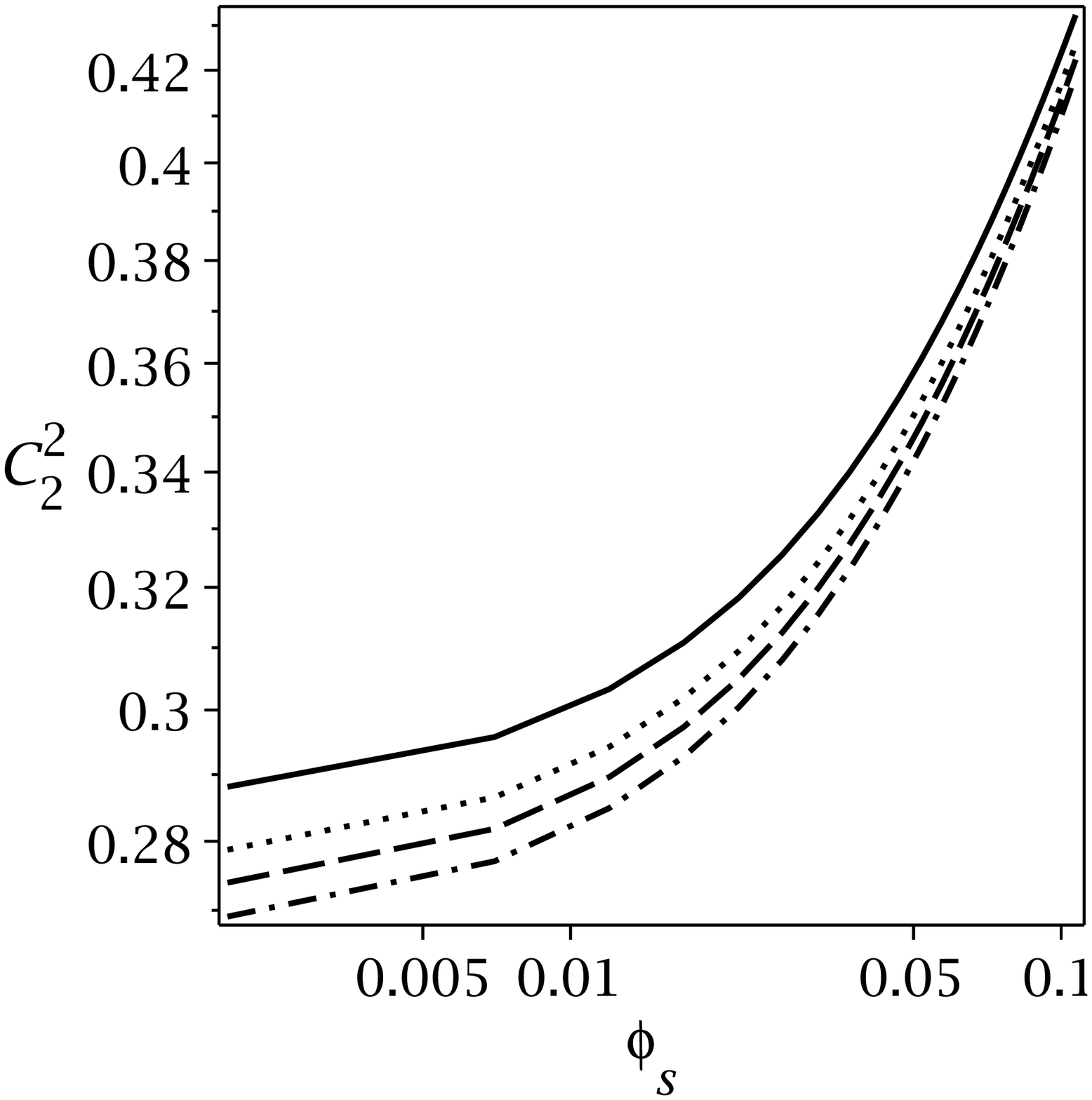}  }
} 
\centerline
{ 
{\epsfxsize=4.3cm\epsffile{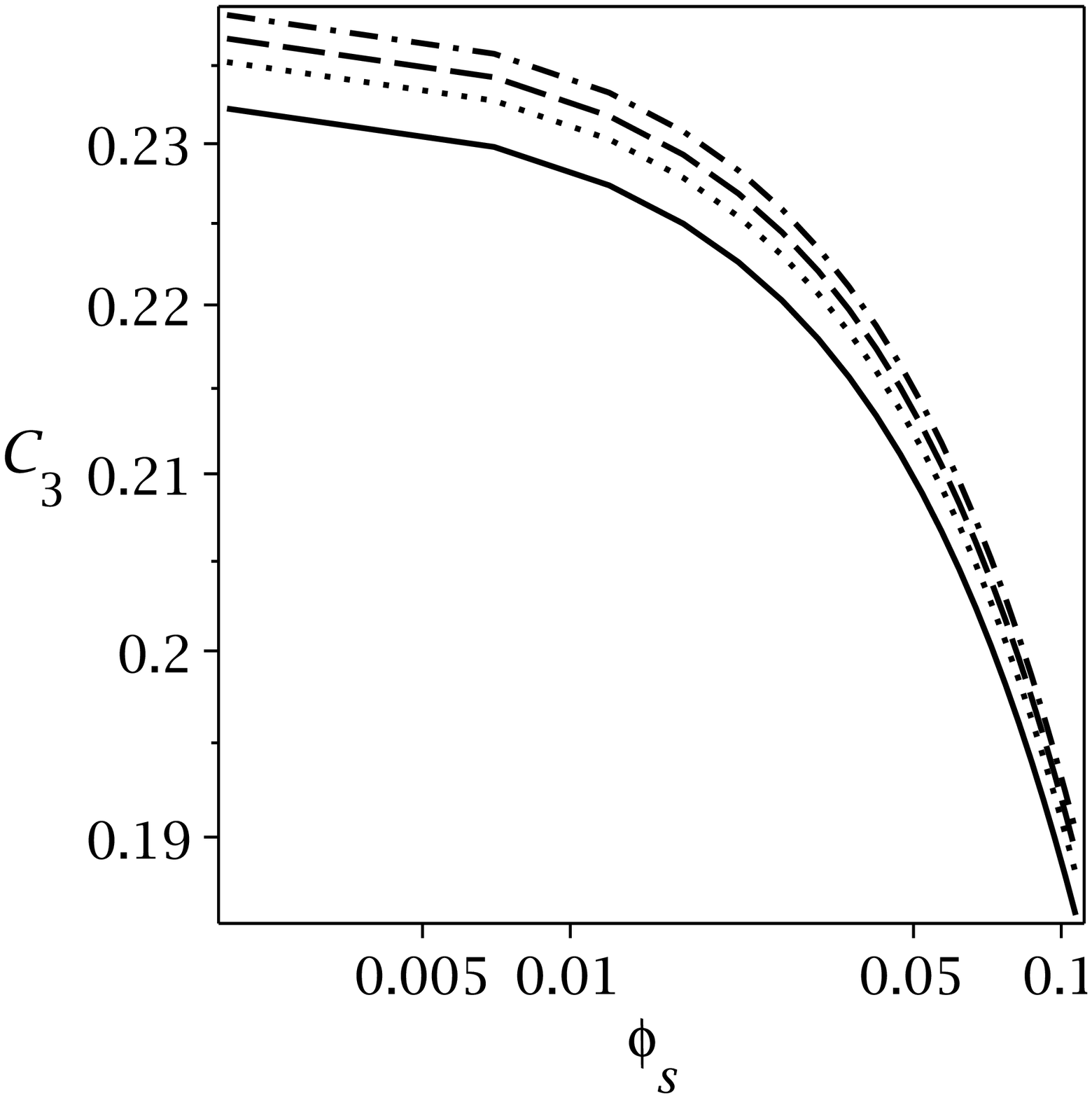}  }
{\epsfxsize=4.3cm\epsffile{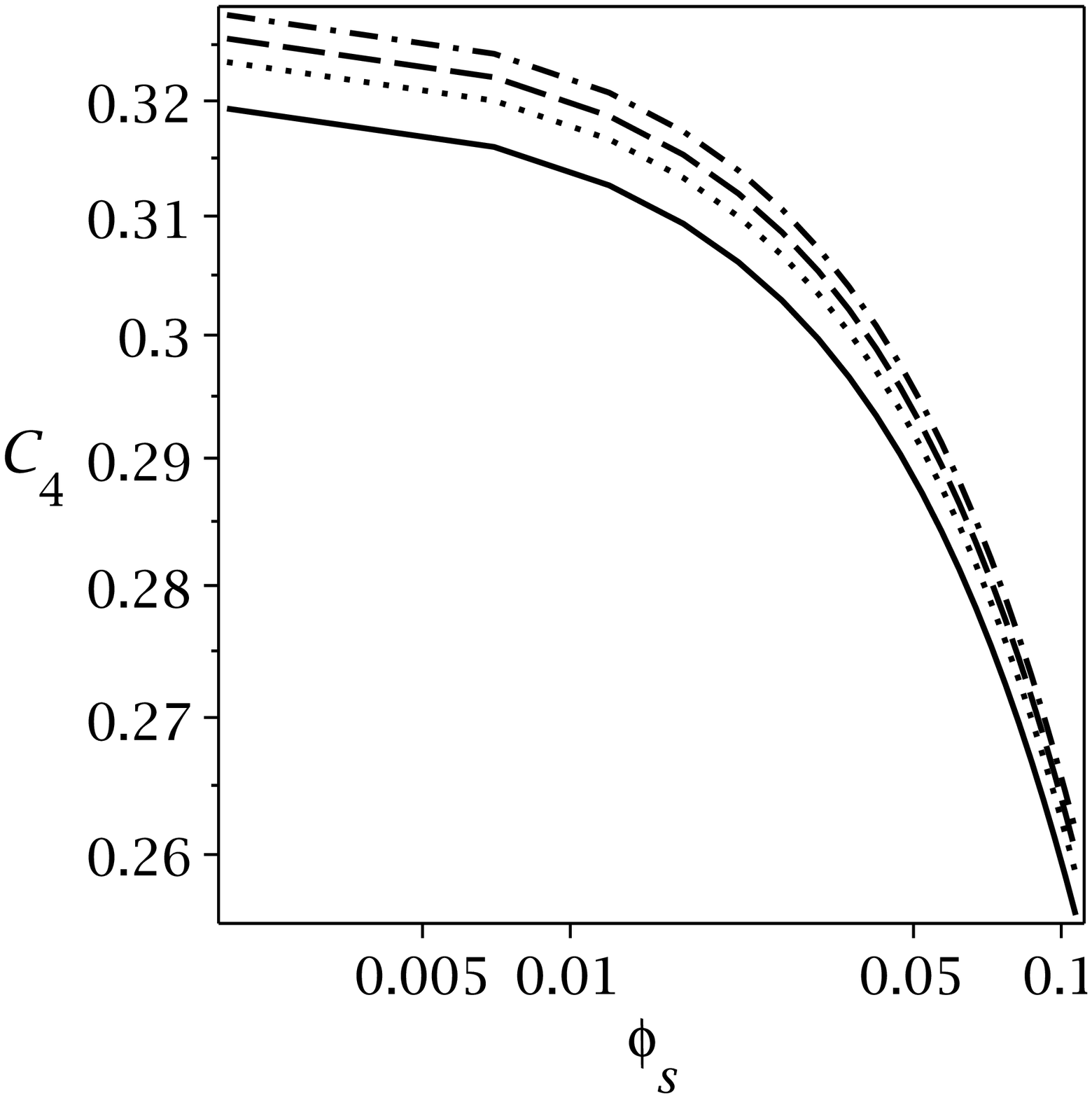}  }
}      
\end{center}
\begin{center}
\vspace{-0.5cm}
\caption{Same as Fig. 1, but $\beta=1.0$, and 
solid, dotted, dashed, and dot-dashed lines represent $P_m = \infty, 1, 2/3$, and $1/2$.}
\end{center}
\end{figure}

\input{epsf}
\begin{figure}[t]
\begin{center}
\centerline
{ 
{\epsfxsize=4.2cm\epsffile{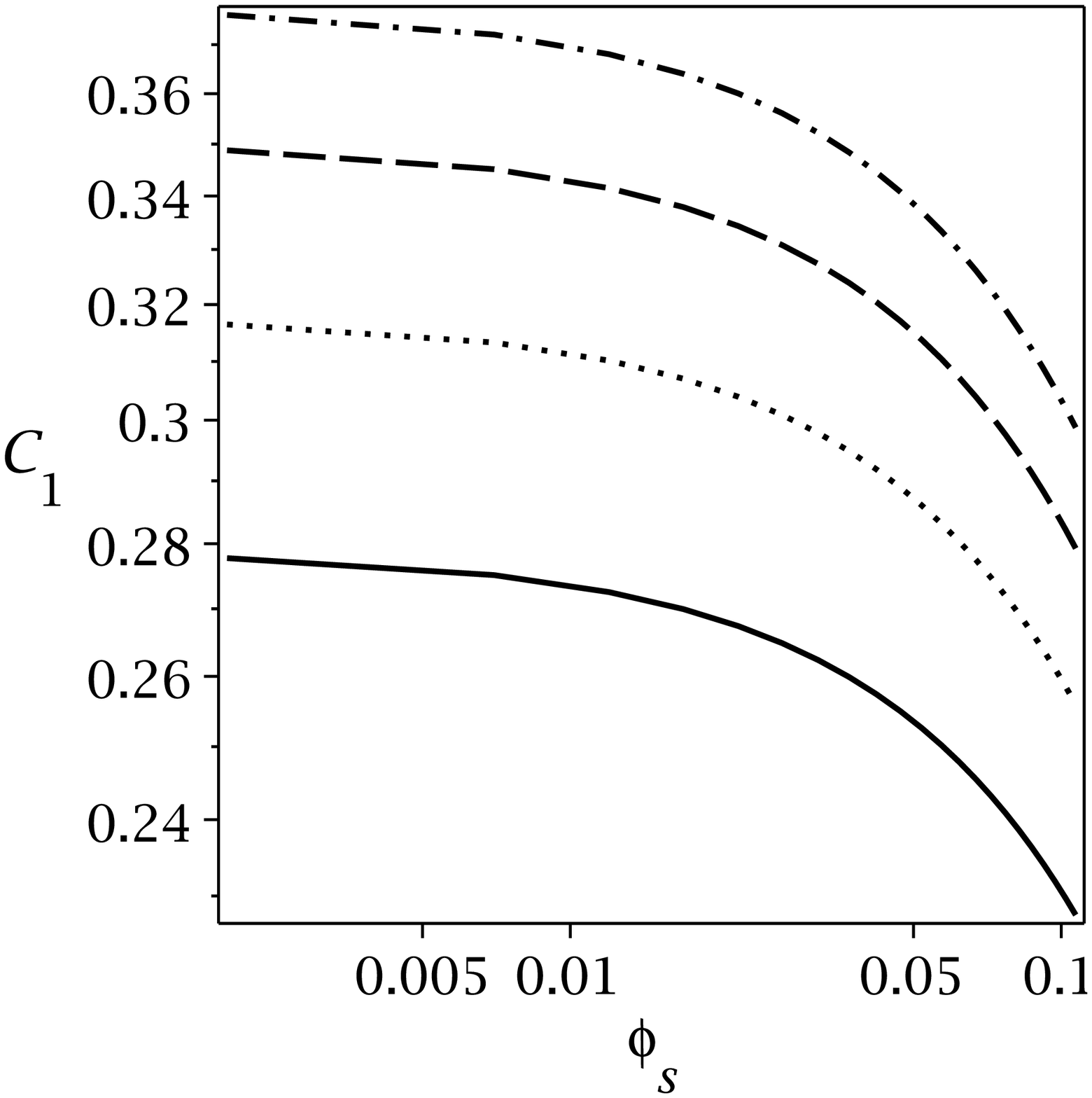}  }
{\epsfxsize=4.2cm\epsffile{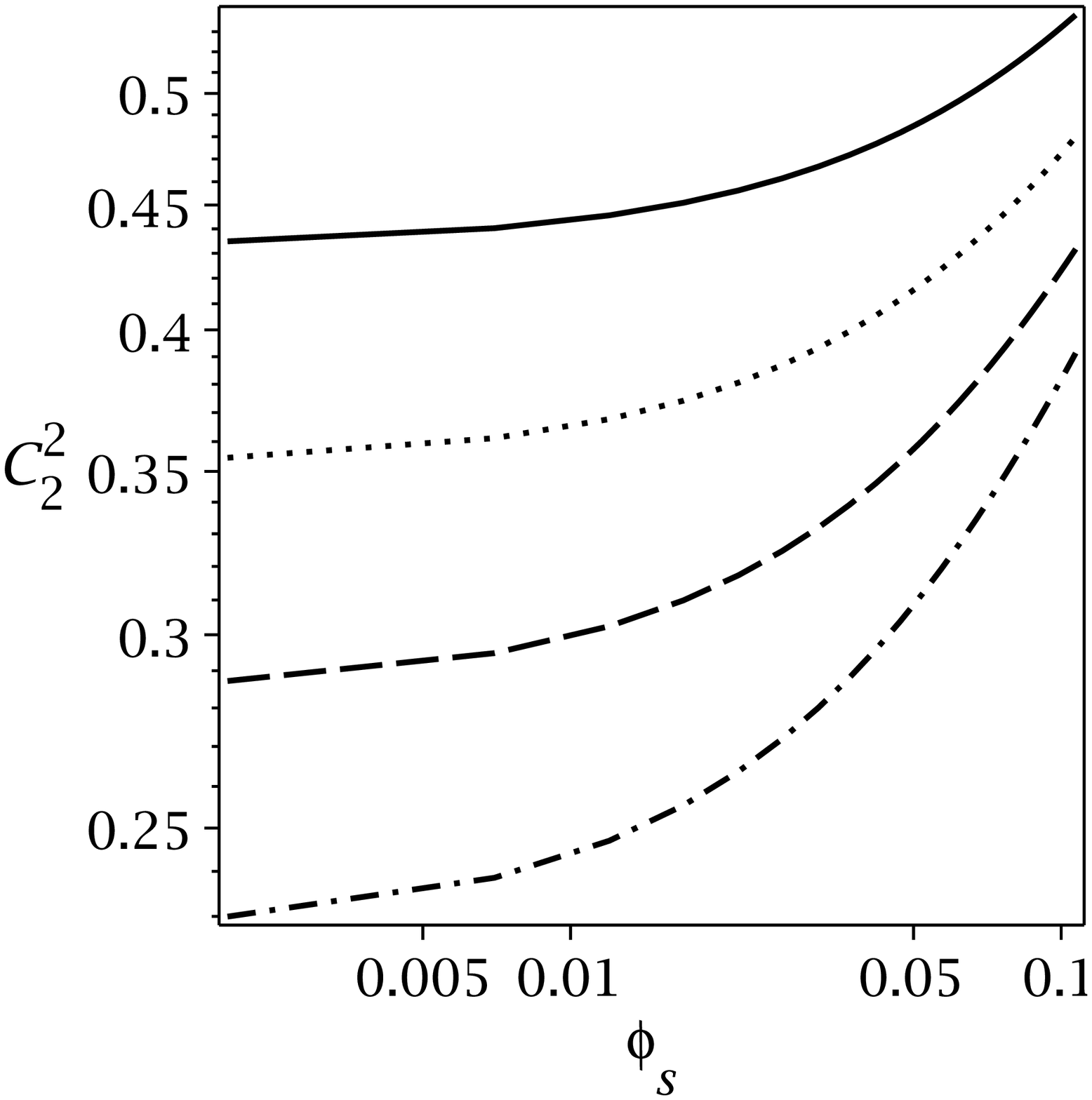}  }
} 
\centerline
{ 
{\epsfxsize=4.2cm\epsffile{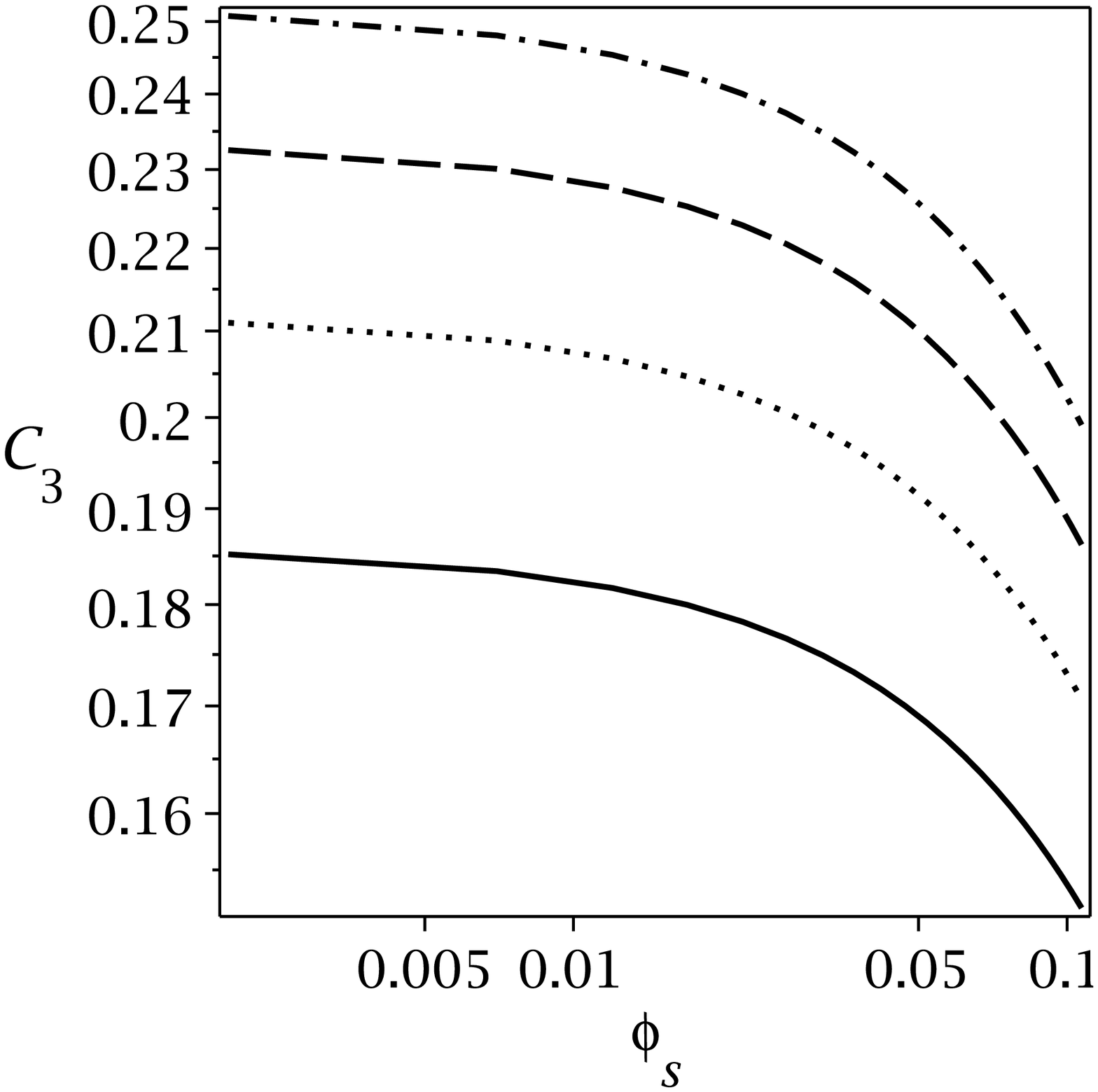}  }
{\epsfxsize=4.2cm\epsffile{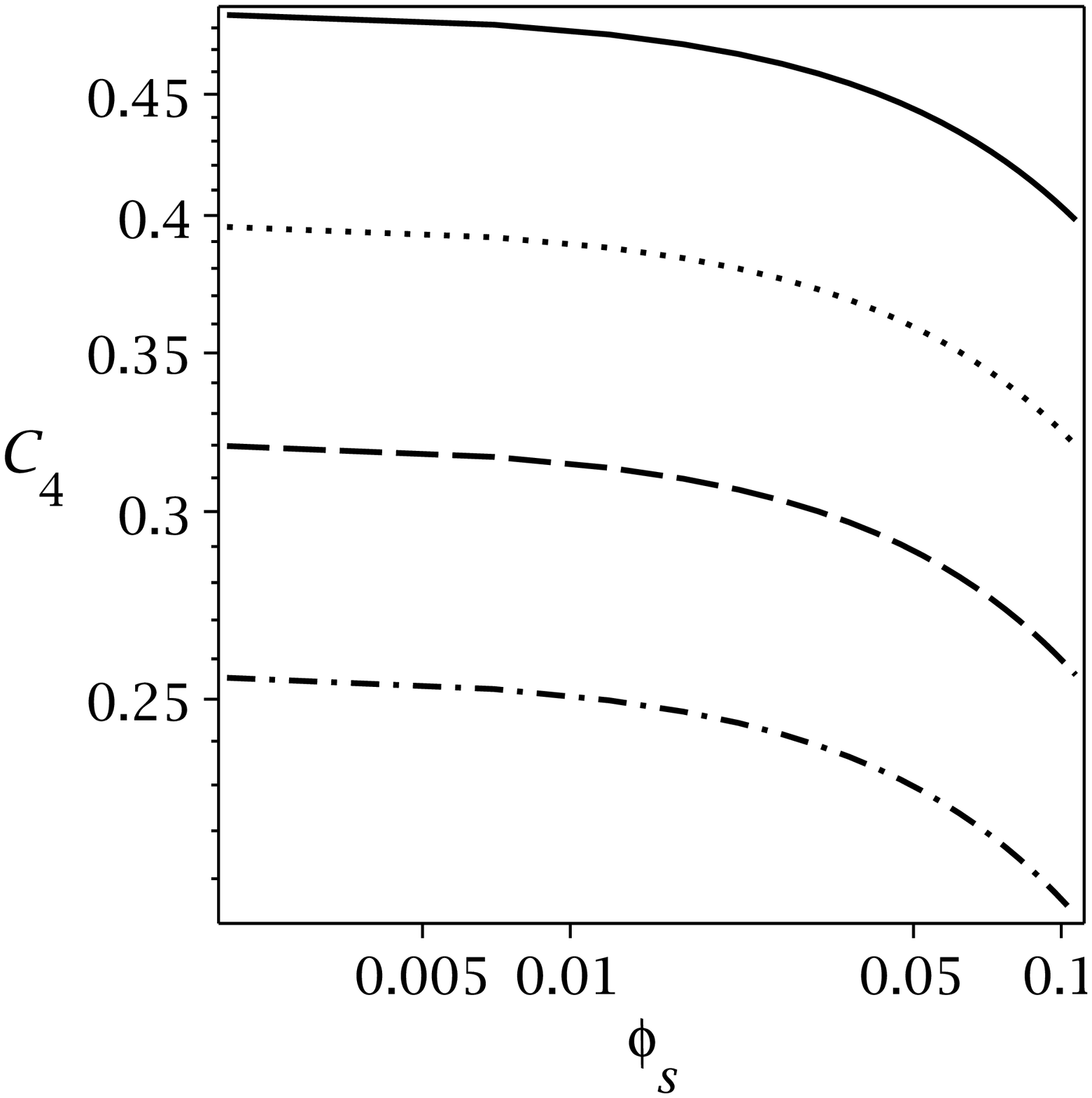}  }
}      
\end{center}
\begin{center}
\vspace{-0.6cm}
\caption{
Same as Fig. 1, but $\beta=1.0$, and 
 solid, dotted, dashed, and dot-dashed lines represent $\gamma = 1.2, 1.25, 1.3$, and $1.35$.
}
\end{center}
\end{figure}

The solutions in Figures 1-3 imply that the radial infall velocity, $c_1$, and the sound speed, $c_3$, both 
decrease with the magnitude of conduction, while the squared angular velocity, $c_2^2$, increases. These properties are 
qualitatively consistent with dynamical analysis of Faghei (2011b). One and two dimensional simulations of hot accretion 
flows have also shown that thermal conduction reduces the flow temperature (Sharmal et al. 2008; Wu et al. 2010). The 
profiles of advection transport of energy, $c_4$, in Figures 1-3 imply that thermal conduction behaves as a cooling 
mechanism, resulting in a local decrease of the gas temperature relative to the original ADAF solution. At the same time, 
the gas adjust its angular velocity (which increases the level of viscous dissipation) and reduces its inflow speed.

Figure 1 shows the radial velocity, sound speed, and advection transport of energy decrease by adding the magnetic pressure 
fraction, $\beta$. These properties are qualitatively consistent with results of Bu et al. (2009) and Faghei (2011a). 
Moreover, the angular velocity decreases with the magnitude of magnetic field that is in according with results of 
Khesali \& Faghei (2009) and Faghei (2011a).

Figure 2 shows the magnetic diffusivity has the opposite effects of thermal conduction on the physical variables. 
As, by adding the magnetic diffusivity, $P_m^{-1}$, the radial velocity, sound speed, and advection transport of energy 
increase, while the rotational velocity decrease. These results are similar to resistive ADAF models without thermal 
conduction (e. g. Faghei 2011a). 

Figure 3 represents the gas adiabatic index similar to magnetic diffusivity has the opposite effects of thermal 
conduction on the physical quantities. As, with the magnitude of $\gamma$, the inflow and sound speed increase, while the 
rotational velocity decreases. This is in accord with dynamical study of hot accretion flow (Faghei 2011b). Moreover, 
Figure 3 shows that the gas adiabatic index contributes with thermal conduction to reduce advection transport of energy. It 
can be due to decrease of rotational velocity by adding $\gamma$, which reduces the level of viscous dissipation. 

The studies of hot accretion flows (e. g. Shadmehri 2008) imply that the solution for a given set of the input parameters 
reaches to a non-rotating limit at a specific of $\phi_s$ which we denote it by $\phi_s^c$. With zero insertion of $c_2$ in 
equations (22)-(24),  $\phi_s^c$ can be written as
\begin{eqnarray}
\nonumber \phi_s^c=\frac{1}{60}\left[\frac{\beta f}{P_m} -  \frac{5/3-\gamma}{\gamma-1} \right] \times
\\ \sqrt{2 (\beta+5) \left[ -1+\sqrt{1+\frac{18 \alpha^2}{(\beta+5)^2}} \,\right]}.
\end{eqnarray} 
We can not extend the studies beyond $\phi_s^c$, because the right-hand side of equation (29) becomes negative and a 
negative $c_2^2$ is clearly unphysical. As, the rotational velocity profiles in Figures 1-3 show, we have selected the 
input parameters that $c_2^2$ is positive. The behaviour of critical saturation constant, $\phi_s^c$, as a function of 
$\beta$ for several values of magnetic Prandtl number is shown in Figures 4. In left panel of Figure 4, the viscosity 
parameter value is $\alpha=0.1$, and in  right panel is $\alpha=0.2$. The profiles of Figure 4 show the critical saturation 
constant highly depends on magnetic pressure fraction, $\beta$, magnetic Prandtl number, $P_m$, and viscosity parameter,
$\alpha$. As, higher values of $\beta$ corresponds to larger $\phi_s^c$. Critical saturation constant also 
increases with higher value of $\alpha$. Since, magnetic diffusivity is proportional to $P_m^{-1}$, Figure 4 shows that the 
magnetic diffusivity similar to magnetic field increases $\phi_s^c$ value.    

As mentioned in the introduction, Sharma et al. (2008) by resistive MHD simulation studied a spherical accretion with 
thermal conduction. They found for even modest thermal conductivities, conduction is the significant mechanism of energy. 
Here, to compare conduction mechanism to others, we study the ratio of energy transport by thermal conduction, $Q_{cond}$, 
to  the gas heating rate by viscosity $Q_{vis}$ and resistivity $Q_{resis}$. Such solutions are shown in Figure 5 for two 
cases of non-resistive (left-panel) and resistive (right-panel) flows. The solutions imply that thermal conduction
is the significant energy mechanism in the flow. This result confirms simulation of Sharma et al. (2008). 
Moreover, the ratio of $Q_{cond}$ to $Q_{diss}$ increases slightly by adding the magnetic field and does not change for 
different values of magnetic Prandtl number. Because, a large fraction of $Q_{diss}$ is generated by viscous dissipation.

\section{Summary and Discussion}
The collision timescale between ions and electrons in hot accretion flows is longer than the inflow timescale. Thus, the 
inflow plasma is collisionless, and transfer of energy by thermal conduction can be dynamically important. The low 
collisional rate of the gas is confirmed by direct observation, particularly in the case of the Galactic centre (Quataert 
2004; Tanaka \& Menou 2006) and in the intracluster medium of galaxy clusters (Sarazin 1986). 

In this paper, the structure of a magnetized ADAF  in the presence of resistivity and thermal conduction is investigated. 
We assumed the magnetic field has a purely toroidal component. We adopted the presented solutions by Tanaka \& Menou (2006) 
and Faghei (2011a). Thus, we assumed that angular momentum transport is due to viscous turbulence and the $\alpha$-
prescription is used for the kinematic coefficient of viscosity. We also assumed the flow does not have a good cooling 
efficiency and so a fraction of energy accretes along with matter on to the central object. In order to solve the equations 
that govern the structure behaviour of magnetized ADAF with thermal conduction, we have used steady self-similar solution.
\input{epsf}
\begin{figure}[t]
\begin{center}
\centerline
{ 
{\epsfxsize=4.3cm\epsffile{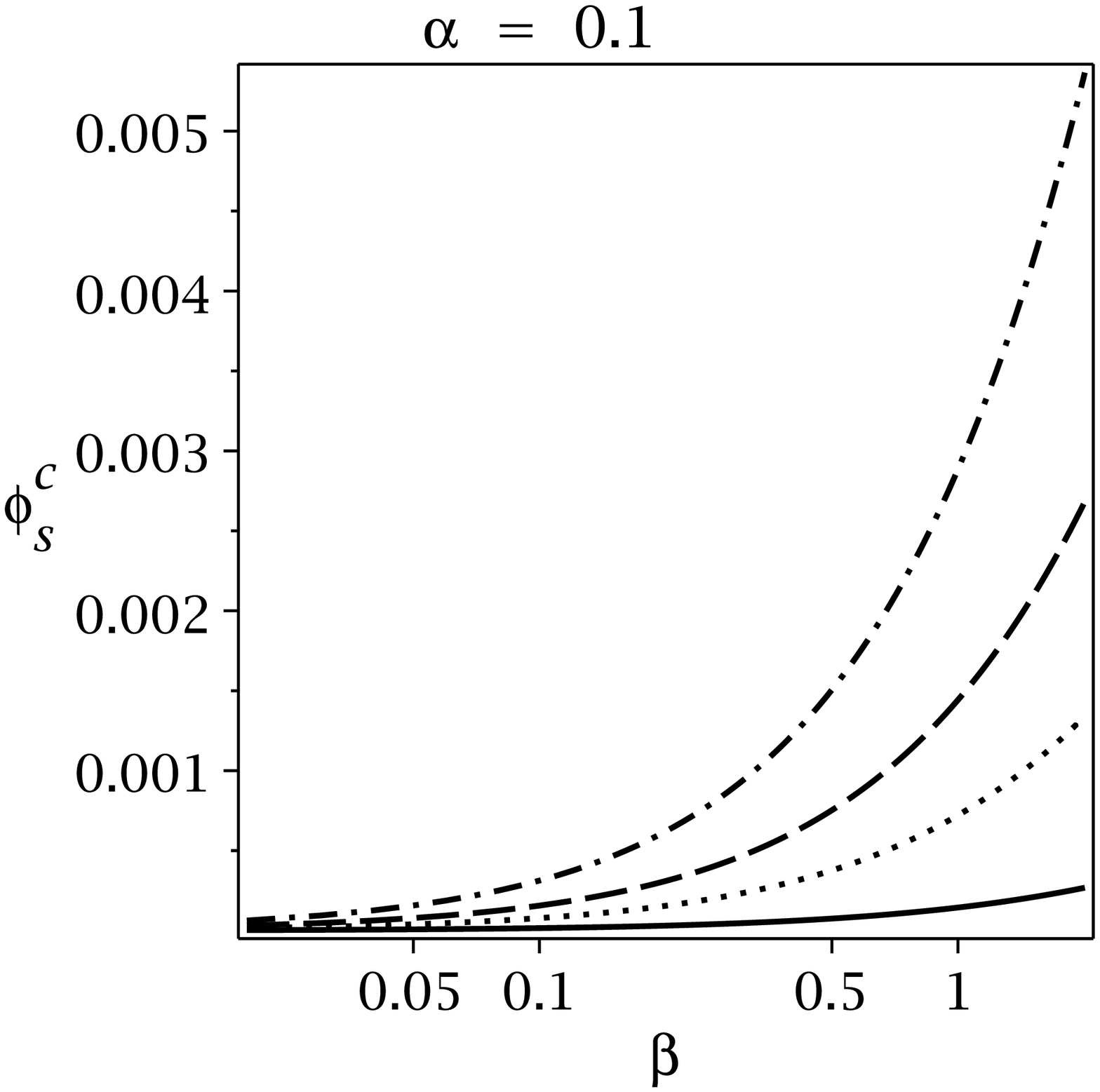}  }
{\epsfxsize=4.3cm\epsffile{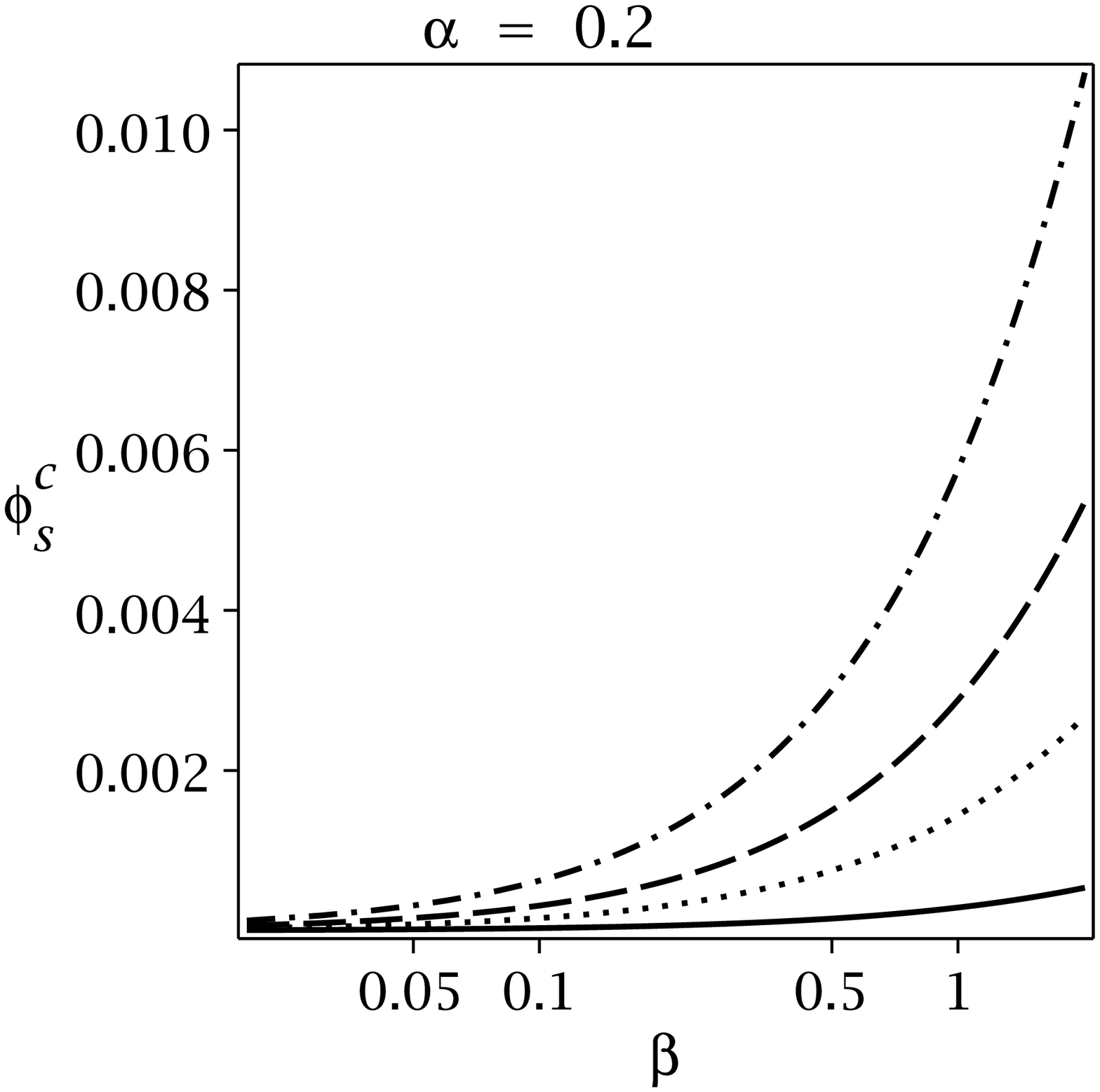}  }
} 
\end{center}
\begin{center}
\caption{
The critical saturation constant as a function of the ratio of magnetic pressure to gas pressure. Solid, dotted, dashed, and dot-dashed lines represent $P_m = 10, 
5, 1$, and $0.5$. The input parameters are set to $\gamma=5/3$, $f=1$, $s=-3/2$, and the viscous parameter $\alpha$ in \textit{left-panel} is $0.1$ and in 
\textit{right-panel} is $0.2$.
}
\end{center}
\end{figure}

The solutions showed the radial infall velocity and sound speed in the presence of thermal conduction both 
decrease, while angular velocity increase. These properties are consistent with dynamical study of hot accretion flow (e. 
g. Faghei 2011b) and from some aspects also are in accord with simulations of Sharma et al. (2008) and Wu et al. (2010). 
Moreover, the solutions represent the magnetic diffusivity and thermal conduction have the opposite effects on physical 
quantities. For a moderate thermal conduction, the solutions imply that thermal conduction can play an important role in 
energy mechanism of the system. This property is qualitatively consistent with non-ideal simulations of Sharma et al. 
(2008). 

In the present model, accretion flow is studied in one-dimensional approach and ignored from latitudinal dependence of 
physical quantities. There are some researches in two-dimensional approach that express the importance of such studies 
(Tanaka \& Menou 2006; Ghanbari et al. 2009; Wu et al. 2010 ). Thus, latitudinal study of present model can be investigated 
in other research. Here, we used a saturated heat conduction flux. While, there are some studies with unsaturated 
heat flux (e. g. Shcherbakkov \& baganoff 2010) that show a good agreement with observations. Thus, the study of the 
present model in a unsaturated case will be interesting.

\input{epsf}
\begin{figure}[t]
\begin{center}
\centerline
{ 
{\epsfxsize=4.3cm\epsffile{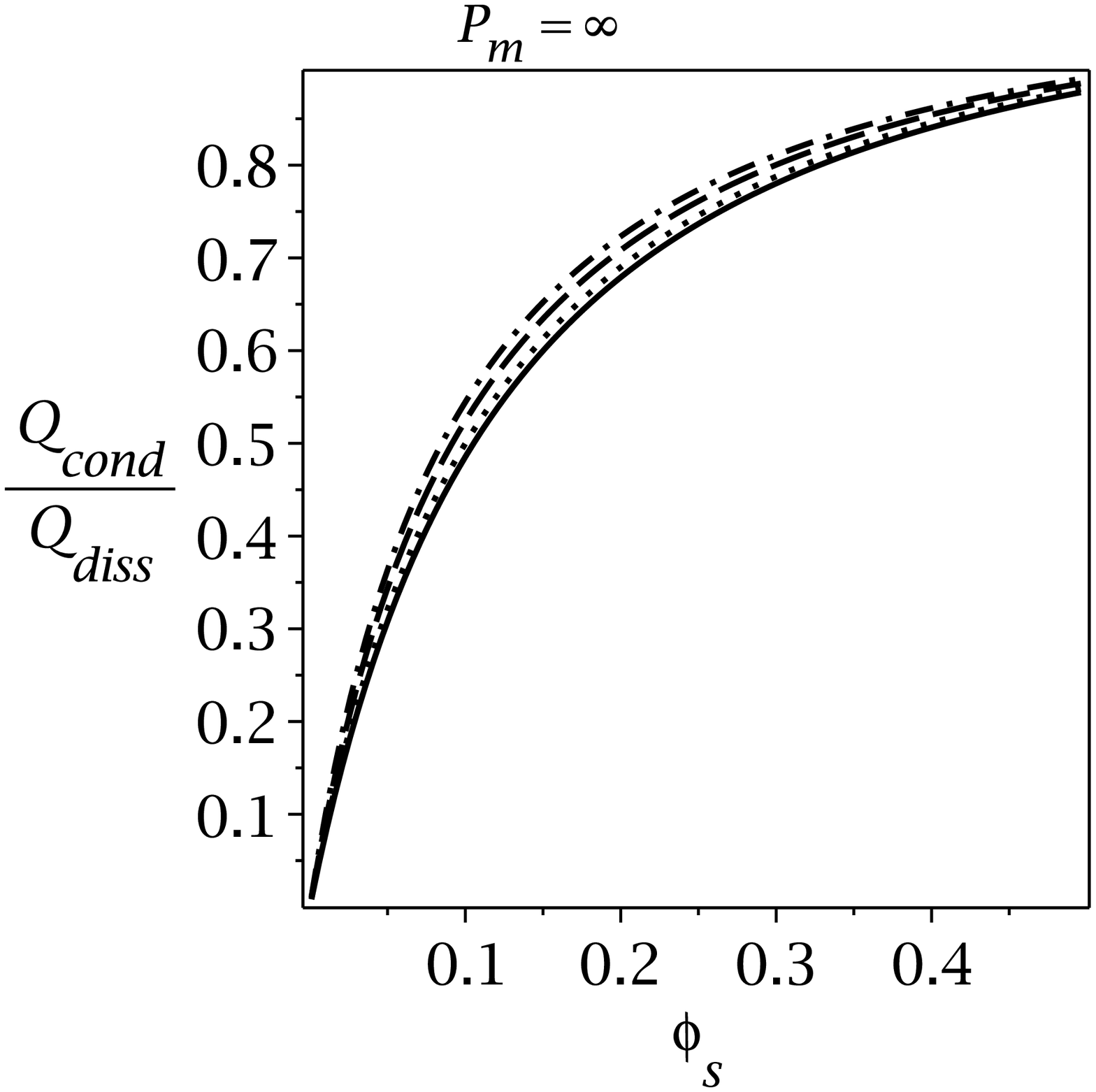}  }
{\epsfxsize=4.3cm\epsffile{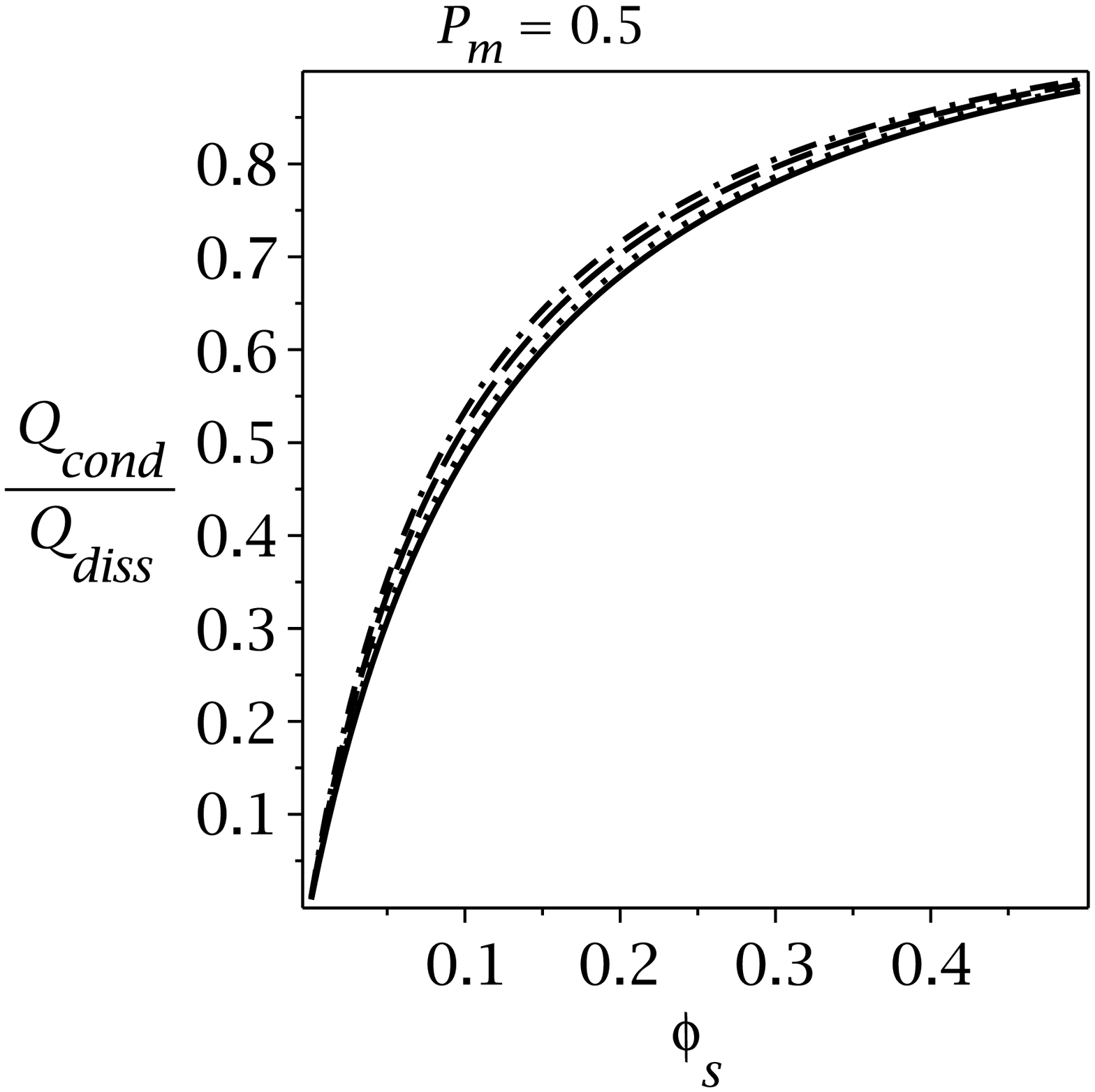}  }
} 
\end{center}
\begin{center}
\caption{
The ratio of energy transport by thermal conduction, $Q_{cond}$, to  the gas heating rate by viscosity $Q_{vis}$ and resistivity $Q_{resis}$ as a function of
 saturation constant. Solid, dotted, dashed, and dot-dashed lines represent $\beta=0.0, 1.0, 3.0$, and $5.0$. The input parameters are set to $\gamma=4/3$, $f=1$,
 $s=-3/2$, $\alpha=0.5$, and the magnetic Prandtl number in \textit{left-panel} is $\infty$ and in \textit{right-panel} is $0.5$.
}
\end{center}
\end{figure}

\section*{Acknowledgements}
I wish to thank the anonymous referee for very useful comments that helped me to improve the initial version of the paper. 
I would also like to thank Roman V. Shcherbakov for his helpful comments.

\end{document}